\documentclass[twocolumn,letterpaper,aps,prl,superscriptaddress,showpacs,floatfix]{revtex4}

\usepackage{graphicx}	
\usepackage{xspace}	

\newcommand{\pt}{\mbox{$p_T$}\xspace}

\newcommand{\Npart}{\mbox{$N_{\rm part}$}\xspace}

\newcommand{\sqsn}{\mbox{$\sqrt{s_{_{NN}}}$}\xspace}

\newcommand{\piz}{\mbox{$\pi^0$}\xspace}
\newcommand{\h}{\mbox{$\eta$}\xspace}
\newcommand{\vtwo}{\mbox{$v_2$}\xspace}
\newcommand{\rgam}{\mbox{$R_{\gamma}$}\xspace}

\begin{document}

\title{Observation of direct-photon collective flow in Au$+$Au collisions 
at $\sqrt{s_{_{NN}}}=200$~GeV}

\newcommand{\abilene}{Abilene Christian University, Abilene, Texas 79699, USA}
\newcommand{\banaras}{Department of Physics, Banaras Hindu University, Varanasi 221005, India}
\newcommand{\barc}{Bhabha Atomic Research Centre, Bombay 400 085, India}
\newcommand{\bnlcoll}{Collider-Accelerator Department, Brookhaven National Laboratory, Upton, New York 11973-5000, USA}
\newcommand{\bnlphys}{Physics Department, Brookhaven National Laboratory, Upton, New York 11973-5000, USA}
\newcommand{\caucr}{University of California - Riverside, Riverside, California 92521, USA}
\newcommand{\charlesczech}{Charles University, Ovocn\'{y} trh 5, Praha 1, 116 36, Prague, Czech Republic}
\newcommand{\chonbuk}{Chonbuk National University, Jeonju, 561-756, Korea}
\newcommand{\ciae}{Science and Technology on Nuclear Data Laboratory, China Institute of Atomic Energy, Beijing 102413, P.~R.~China}
\newcommand{\cns}{Center for Nuclear Study, Graduate School of Science, University of Tokyo, 7-3-1 Hongo, Bunkyo, Tokyo 113-0033, Japan}
\newcommand{\colorado}{University of Colorado, Boulder, Colorado 80309, USA}
\newcommand{\columbia}{Columbia University, New York, New York 10027 and Nevis Laboratories, Irvington, New York 10533, USA}
\newcommand{\czechtech}{Czech Technical University, Zikova 4, 166 36 Prague 6, Czech Republic}
\newcommand{\dapnia}{Dapnia, CEA Saclay, F-91191, Gif-sur-Yvette, France}
\newcommand{\debrecen}{Debrecen University, H-4010 Debrecen, Egyetem t{\'e}r 1, Hungary}
\newcommand{\elte}{ELTE, E{\"o}tv{\"o}s Lor{\'a}nd University, H - 1117 Budapest, P{\'a}zm{\'a}ny P. s. 1/A, Hungary}
\newcommand{\ewha}{Ewha Womans University, Seoul 120-750, Korea}
\newcommand{\fit}{Florida Institute of Technology, Melbourne, Florida 32901, USA}
\newcommand{\fsu}{Florida State University, Tallahassee, Florida 32306, USA}
\newcommand{\gsu}{Georgia State University, Atlanta, Georgia 30303, USA}
\newcommand{\hiroshima}{Hiroshima University, Kagamiyama, Higashi-Hiroshima 739-8526, Japan}
\newcommand{\ihepprot}{IHEP Protvino, State Research Center of Russian Federation, Institute for High Energy Physics, Protvino, 142281, Russia}
\newcommand{\illuiuc}{University of Illinois at Urbana-Champaign, Urbana, Illinois 61801, USA}
\newcommand{\inrras}{Institute for Nuclear Research of the Russian Academy of Sciences, prospekt 60-letiya Oktyabrya 7a, Moscow 117312, Russia}
\newcommand{\instpasczech}{Institute of Physics, Academy of Sciences of the Czech Republic, Na Slovance 2, 182 21 Prague 8, Czech Republic}
\newcommand{\isu}{Iowa State University, Ames, Iowa 50011, USA}
\newcommand{\jinrdubna}{Joint Institute for Nuclear Research, 141980 Dubna, Moscow Region, Russia}
\newcommand{\jyvaskyla}{Helsinki Institute of Physics and University of Jyv{\"a}skyl{\"a}, P.O.Box 35, FI-40014 Jyv{\"a}skyl{\"a}, Finland}
\newcommand{\kek}{KEK, High Energy Accelerator Research Organization, Tsukuba, Ibaraki 305-0801, Japan}
\newcommand{\kfki}{KFKI Research Institute for Particle and Nuclear Physics of the Hungarian Academy of Sciences (MTA KFKI RMKI), H-1525 Budapest 114, POBox 49, Budapest, Hungary}
\newcommand{\korea}{Korea University, Seoul, 136-701, Korea}
\newcommand{\kurchatov}{Russian Research Center ``Kurchatov Institute", Moscow, 123098 Russia}
\newcommand{\kyoto}{Kyoto University, Kyoto 606-8502, Japan}
\newcommand{\labllr}{Laboratoire Leprince-Ringuet, Ecole Polytechnique, CNRS-IN2P3, Route de Saclay, F-91128, Palaiseau, France}
\newcommand{\lawllnl}{Lawrence Livermore National Laboratory, Livermore, California 94550, USA}
\newcommand{\losalamos}{Los Alamos National Laboratory, Los Alamos, New Mexico 87545, USA}
\newcommand{\lpc}{LPC, Universit{\'e} Blaise Pascal, CNRS-IN2P3, Clermont-Fd, 63177 Aubiere Cedex, France}
\newcommand{\lund}{Department of Physics, Lund University, Box 118, SE-221 00 Lund, Sweden}
\newcommand{\maryland}{University of Maryland, College Park, Maryland 20742, USA}
\newcommand{\mass}{Department of Physics, University of Massachusetts, Amherst, Massachusetts 01003-9337, USA }
\newcommand{\muenster}{Institut fur Kernphysik, University of Muenster, D-48149 Muenster, Germany}
\newcommand{\muhlenberg}{Muhlenberg College, Allentown, Pennsylvania 18104-5586, USA}
\newcommand{\myongji}{Myongji University, Yongin, Kyonggido 449-728, Korea}
\newcommand{\nagasaki}{Nagasaki Institute of Applied Science, Nagasaki-shi, Nagasaki 851-0193, Japan}
\newcommand{\newmex}{University of New Mexico, Albuquerque, New Mexico 87131, USA }
\newcommand{\nmsu}{New Mexico State University, Las Cruces, New Mexico 88003, USA}
\newcommand{\ornl}{Oak Ridge National Laboratory, Oak Ridge, Tennessee 37831, USA}
\newcommand{\orsay}{IPN-Orsay, Universite Paris Sud, CNRS-IN2P3, BP1, F-91406, Orsay, France}
\newcommand{\peking}{Peking University, Beijing 100871, P.~R.~China}
\newcommand{\pnpi}{PNPI, Petersburg Nuclear Physics Institute, Gatchina, Leningrad region 188300, Russia}
\newcommand{\riken}{RIKEN Nishina Center for Accelerator-Based Science, Wako, Saitama 351-0198, Japan}
\newcommand{\rikjrbrc}{RIKEN BNL Research Center, Brookhaven National Laboratory, Upton, New York 11973-5000, USA}
\newcommand{\rikkyo}{Physics Department, Rikkyo University, 3-34-1 Nishi-Ikebukuro, Toshima, Tokyo 171-8501, Japan}
\newcommand{\saispbstu}{Saint Petersburg State Polytechnic University, St. Petersburg, 195251 Russia}
\newcommand{\saopaulo}{Universidade de S{\~a}o Paulo, Instituto de F\'{\i}sica, Caixa Postal 66318, S{\~a}o Paulo CEP05315-970, Brazil}
\newcommand{\seoulnat}{Seoul National University, Seoul, Korea}
\newcommand{\stonybrkc}{Chemistry Department, Stony Brook University, SUNY, Stony Brook, New York 11794-3400, USA}
\newcommand{\stonycrkp}{Department of Physics and Astronomy, Stony Brook University, SUNY, Stony Brook, New York 11794-3400, USA}
\newcommand{\tenn}{University of Tennessee, Knoxville, Tennessee 37996, USA}
\newcommand{\titech}{Department of Physics, Tokyo Institute of Technology, Oh-okayama, Meguro, Tokyo 152-8551, Japan}
\newcommand{\tsukuba}{Institute of Physics, University of Tsukuba, Tsukuba, Ibaraki 305, Japan}
\newcommand{\vandy}{Vanderbilt University, Nashville, Tennessee 37235, USA}
\newcommand{\waseda}{Waseda University, Advanced Research Institute for Science and Engineering, 17 Kikui-cho, Shinjuku-ku, Tokyo 162-0044, Japan}
\newcommand{\weizmann}{Weizmann Institute, Rehovot 76100, Israel}
\newcommand{\yonsei}{Yonsei University, IPAP, Seoul 120-749, Korea}
\affiliation{\abilene}
\affiliation{\banaras}
\affiliation{\barc}
\affiliation{\bnlcoll}
\affiliation{\bnlphys}
\affiliation{\caucr}
\affiliation{\charlesczech}
\affiliation{\chonbuk}
\affiliation{\ciae}
\affiliation{\cns}
\affiliation{\colorado}
\affiliation{\columbia}
\affiliation{\czechtech}
\affiliation{\dapnia}
\affiliation{\debrecen}
\affiliation{\elte}
\affiliation{\ewha}
\affiliation{\fit}
\affiliation{\fsu}
\affiliation{\gsu}
\affiliation{\hiroshima}
\affiliation{\ihepprot}
\affiliation{\illuiuc}
\affiliation{\inrras}
\affiliation{\instpasczech}
\affiliation{\isu}
\affiliation{\jinrdubna}
\affiliation{\jyvaskyla}
\affiliation{\kek}
\affiliation{\kfki}
\affiliation{\korea}
\affiliation{\kurchatov}
\affiliation{\kyoto}
\affiliation{\labllr}
\affiliation{\lawllnl}
\affiliation{\losalamos}
\affiliation{\lpc}
\affiliation{\lund}
\affiliation{\maryland}
\affiliation{\mass}
\affiliation{\muenster}
\affiliation{\muhlenberg}
\affiliation{\myongji}
\affiliation{\nagasaki}
\affiliation{\newmex}
\affiliation{\nmsu}
\affiliation{\ornl}
\affiliation{\orsay}
\affiliation{\peking}
\affiliation{\pnpi}
\affiliation{\riken}
\affiliation{\rikjrbrc}
\affiliation{\rikkyo}
\affiliation{\saispbstu}
\affiliation{\saopaulo}
\affiliation{\seoulnat}
\affiliation{\stonybrkc}
\affiliation{\stonycrkp}
\affiliation{\tenn}
\affiliation{\titech}
\affiliation{\tsukuba}
\affiliation{\vandy}
\affiliation{\waseda}
\affiliation{\weizmann}
\affiliation{\yonsei}
\author{A.~Adare} \affiliation{\colorado}
\author{S.~Afanasiev} \affiliation{\jinrdubna}
\author{C.~Aidala} \affiliation{\mass}
\author{N.N.~Ajitanand} \affiliation{\stonybrkc}
\author{Y.~Akiba} \affiliation{\riken} \affiliation{\rikjrbrc}
\author{H.~Al-Bataineh} \affiliation{\nmsu}
\author{J.~Alexander} \affiliation{\stonybrkc}
\author{K.~Aoki} \affiliation{\kyoto} \affiliation{\riken}
\author{Y.~Aramaki} \affiliation{\cns}
\author{E.T.~Atomssa} \affiliation{\labllr}
% ...
\collaboration{PHENIX Collaboration} \noaffiliation

%
%

%

%

\date{\today}

\begin{abstract}

The second Fourier component $v_2$ of the azimuthal 
anisotropy with respect to the reaction plane was measured for 
direct photons at midrapidity and transverse momentum ($p_T$) of 
1--13~GeV/$c$ in Au$+$Au collisions at $\sqrt{s_{_{NN}}}=200$~GeV.  
Previous measurements of this quantity for hadrons with $p_T<6$~GeV/$c$ 
indicate that the medium behaves like a nearly perfect fluid, while for 
$p_T>6$~GeV/$c$ a reduced anisotropy is interpreted in terms of a 
path-length dependence for parton energy loss.  In this measurement 
with the PHENIX detector at the Relativistic Heavy Ion Collider we 
find that for $p_T>4$~GeV/$c$ the anisotropy for direct photons is 
consistent with zero, as expected if the dominant source of direct
photons is initial hard scattering.  However, in the $p_T<4$~GeV/$c$ 
region dominated by thermal photons, we find a substantial direct photon 
$v_2$ comparable to that of hadrons, whereas model calculations for 
thermal photons in this kinematic region significantly underpredict 
the observed $v_2$.

\end{abstract}

	



\maketitle


Direct photons are produced in various processes during the entire 
space-time history of relativistic heavy ion collisions
and, due to their small coupling, can leave the collision region
without appreciable 
further interaction.  This makes them a sensitive and direct probe 
of all stages of the collision, including 
initial hard scattering, formation and evolution of the
strongly interacting partonic medium, its transition to hadronic
matter, and final decoupling~\cite{trg2004,liufries}.  
The transverse momentum (\pt) ranges
populated by various production mechanisms overlap.
However, azimuthal asymmetries tied to the event-by-event collision
geometry provide useful additional information and a means to
distinguish between sources of direct photons.  In this paper we
consider the second Fourier component (\vtwo, often referred to as elliptic
flow) of the
event-by-event photon distribution in azimuth with respect to the
reaction plane for minimum bias
and selected centralities in Au$+$Au collisions.

At higher \pt ($>4$~GeV/$c$) there are four
fundamental sources of direct photons, characterized by
different \vtwo~\cite{turbide,liufries}.  Photons from initial hard scattering
(predominantly from $qg\rightarrow q\gamma$ ``gluon Compton
scattering'') are isotropic and so $v_2 = 0$.  
Jet fragmentation photons
have positive \vtwo since the energy loss of the originating parton is
smaller in the reaction plane~\cite{ppg092}.
Jet-conversion photons 
where a hard scattered quark interacts with a
thermal gluon in the medium and converts into a photon with almost
equal \pt  have negative \vtwo~\cite{turbide}, 
because the average pathlength of the
parton in the medium (proportional to the conversion probability)
is larger out of the reaction plane than
within.  Finally, Bremsstrahlung photons are also emitted
preferentially in the direction where the medium is thicker, leading
to a negative \vtwo~\cite{turbide}.
Note that in this picture the azimuthal asymmetry of high \pt
photon production -- while expressed in terms of \vtwo  --
reflects the pure geometry of the medium, not its dynamics: it
depends on the pathlength, not on the boost from the hydrodynamic
pressure gradients.  

The picture is quite different in the low \pt range 
($1<p_T<4$~GeV/$c$) 
dominated by thermal photons, as first measured in~\cite{ppg086}, where
bulk dynamics (expansion) plays an important role since it influences 
both the rate and azimuthal asymmetries of photon
production~\cite{turbide,dusling}.  It is
now established that collectivity -- which already exists in the
partonic phase (strongly interacting Quark-Gluon Plasma, sQGP) --
persists after transition into the hadronic phase and the resulting
azimuthal asymmetries in particle production can be described by
near-ideal hydrodynamics.  The expectation is that thermal
radiation from both the sQGP and the hadronic phase will 
inherit the collective motion of the medium, i.e. will have
a bona fide elliptic flow, positive \vtwo at low \pt~\cite{gale}. 
The low \pt behavior
of direct photon \vtwo puts constraints on the viscosity of the
sQGP~\cite{dusling}.

The PHENIX experiment has published the invariant yield as a function of \pt
for direct photons both via real photons and internal conversions of nearly
real virtual photons~\cite{ppg042,ppg086}.  
In the $1<p_T<4$~GeV/$c$ region, a substantial excess 
of direct photons was observed relative to scaling of $p+p$ 
yields and has been interpreted in terms of thermal photon emission
from the hot medium.
An early attempt to infer \vtwo of direct photons from a \piz and
inclusive photon \vtwo measurement performed in a limited \pt range 
has been published in~\cite{ppg046}.  
In this Letter we present measurements by the PHENIX
experiment~\cite{phenix} of \vtwo of
\piz and inclusive photons in a much extended transverse momentum
(\pt) range (up to 13~GeV/$c$)
in \sqsn = 200~GeV Au$+$Au collisions.  Also, at low \pt
the fraction $R_\gamma$ of direct over inclusive photons is now
measured with much higher precision~\cite{ppg086} than
before~\cite{ppg042}, therefore, for the first time a meaningful
extraction of the direct photon \vtwo itself is possible.

Data were taken in the 2007 run of the Relativistic Heavy Ion
Collider at Brookhaven National Laboratory.  
The analyzed sample 
includes $\sim 3.0 \times10^9$ minimum bias Au$+$Au collisions.
Events were triggered by the Beam-Beam Counters (BBC), as described 
in~\cite{bbc}, which comprise two
arrays of \v{C}erenkov counters covering $3.1<|\eta|<3.9$ and $2\pi$ in
azimuth in both beam directions (North and South).  Event centrality
was determined by the charge sum in the BBC.

The event-by-event reaction plane (RP) has been determined by two 
detectors, the first being the BBC itself.
The RP resolution 
(effectively a dilution factor with which the observed \vtwo 
is normalized to obtain the true \vtwo) is defined as
$\sigma_{\rm RP}=<\cos[2(\Psi^{\rm true}-\Psi^{\rm RP})]>$
and it is established by comparing event-by-event the RPs obtained 
separately in the North and South detectors.
The resolution is highest in the 20-30\% centrality bin where 
it reaches a value of 0.4.  For
the 2007 data taking period, a dedicated reaction plane detector
(RXN)~\cite{rxnp} 
was installed covering $1.0<|\eta|<2.8$ and
the full azimuth.  The RXN is a highly segmented lead-scintillator
sampling detector providing much better measurement 
($\sigma_{\rm RP}\sim$0.7) than
the BBC, but it is closer to the central $|\eta|<0.35$ pseudorapidity 
region where \vtwo is measured, making it more sensitive to jet bias
in those (rare) events where a high \pt particle is observed.  
The $0.7/0.4=1.75$ improvement on the
reaction plane resolution results is a 1.75-fold improvement on 
point-by-point uncertainty. 

Inclusive photons were measured in the
PHENIX electromagnetic calorimeter~\cite{emc}.
Particles were identified (PID) and hadrons were rejected by a
shower shape cut and a veto on charged particles 
using the Pad Chambers~\cite{pc}.  
The remaining sample is collected for each \pt range in histograms 
binned according to $\Phi - \Psi^{\rm RP}$ where
$\Psi^{\rm RP}$ is the azimuth of the event-by-event reaction
plane 
and established independently by the BBC and RXN.  
These distributions are then fit for each \pt range with 
$N_0~[1~+~2~v_2~\cos\{2(\Phi-\Psi^{\rm RP})\}]$
to extract the raw $v_2^{\gamma,{\rm meas}}$ coefficient 
for inclusive photons.  As a cross-check of the fit value, 
another $v_2^{\gamma,{\rm meas}}$ is also calculated from the average
cosine of the particles with respect to the reaction plane.
While the PID eliminates virtually all hadrons 
above 6~GeV deposited energy (which might come
from hadrons of {\it any} \pt above 6~GeV/$c$), 
a significant fraction of hadrons (up to 20\% below 2~GeV deposited energy) 
survive the photon identification cuts.
Since hadrons are known to have a large \vtwo value,
the observed $v_2^{\rm obs}$ of inclusive photons is obtained
after correcting for hadrons as

\begin{displaymath}
 v_2^{\gamma,{\rm obs}} =
  \frac{v_2^{\gamma,{\rm meas}} - (N^{\rm hadr}/N^{\rm meas}) v_2^{\rm hadr}}
       {1 - N^{\rm hadr}/N^{\rm meas}},
\end{displaymath}

\noindent
where $v_2^{\rm hadr}$ is the elliptic flow of hadrons
and $N^{\rm hadr}/N^{\rm meas}$ is the fraction of hadrons in the sample 
surviving the PID cuts, as estimated from {\sc geant} simulations 
(20\% at 2~GeV, 10\% at 4~GeV and negligible above  6~GeV 
deposited energy).
Finally the true $v_2^{\gamma,{\rm inc}}$ for inclusive photons is obtained by 
dividing by the reaction plane resolution
$v_2^{\gamma,{\rm inc}} = v_2^{\gamma,{\rm obs}}/{\sigma_{\rm RP}}$.

A large fraction of
inclusive photons comes from hadron decays, predominantly from \piz
($\sim$80\%) and \h ($\sim$15\%), 
with a small fraction coming from $\rho, \omega$ and $\eta'$
decays, but only the \piz \vtwo is directly measured.  
The measurement of neutral pions and their \vtwo is described in
detail in~\cite{ppg092,ppg110}.
We assume that $\eta$, $\omega$, etc. follow the same $KE_T$ scaling 
observed in hadrons~\cite{ppg062} 
where $KE_T = m_T - m$, 
Thus, $v_2^{\rm hadr}(p_T)$ can be calculated for all hadrons
from $v_2^{\pi^0}(p_T)$.
For this we assume $m_T$-scaling of hadron \pt spectra and
establish a ``hadron cocktail'' using the measured yield ratios,
similar to the one in~\cite{ppg086}.
This cocktail is the input
of a Monte Carlo simulation to calculate the total
$v_2^{\gamma,{\rm bg}}$ due to photons from hadron decays.  
The direct photon $v_2^{\gamma,{\rm dir}}$ 
is then obtained using the $R_{\gamma}(p_T)$ ``direct
photon excess ratio'' as

\begin{displaymath}
v_2^{\gamma,{\rm dir}} =
\frac{R_{\gamma}(p_T) v_2^{\gamma,{\rm inc}} - v_2^{\gamma,{\rm bg}}}
{R_{\gamma}(p_T) - 1},
\end{displaymath}

\noindent
where $R_{\gamma}(p_T)=N^{\rm inc}(p_T)/N^{bg}(p_T)$ with
$N^{\rm inc}=N^{\rm meas}-N^{\rm hadr}$, the number of inclusive
photons, while $N^{bg}(p_T)$ is the number of photons attributed to
hadron decay.  Values of
$R_{\gamma}(p_T)$ above 5~GeV/$c$ are taken from the real photon 
measurement
with the PHENIX electromagnetic calorimeter~\cite{ppg042}, and below that 
from the more accurate, but \pt-range limited internal conversion 
measurement of direct photons~\cite{ppg086}.  

\begingroup \squeezetable

\begin{table}[tbh]
\caption{\label{tab:syserr}
Systematic uncertainties ($\delta x/x$) contributing to the direct photon 
$v_2^{\gamma,{\rm dir}}$ measurement for minimum-bias collisions 
over two \pt ranges.
 }
\begin{ruledtabular}
 \begin{tabular}{ccccc}
  Contributing & Source  & \multicolumn{2}{c}{\pt range} & Type \\ 
  via & & 1-3~GeV/$c$ &  10-16~GeV/$c$ & \\ \hline
   $v_2^{\gamma,{\rm inc}}$ 
   & remaining hadrons  & 2.2\%    &  N/A  &  B    \\
   & \vtwo extraction method  & 0.4\%    & 0.6\%   &  B    \\ \\
   $v_2^{\pi^0}$
   & particle ID  &  3.7\%   &  6.0\%  &  B    \\
   & normalization  &  0.4\%   &  7.2\%  &  B    \\
   & shower merging  &  N/A   &  4.0\%  &  B    \\ \\
   subtraction 
   & $R_{\gamma}$  &  3.1\%   &  22\%  &  B    \\ \\
   common 
   & reaction plane  &  6.3\%   &  6.3\%  &  C    \\
 \end{tabular}
 \end{ruledtabular}
\end{table}

\endgroup

\begin{figure}[thb]
\includegraphics[width=1.0\linewidth]{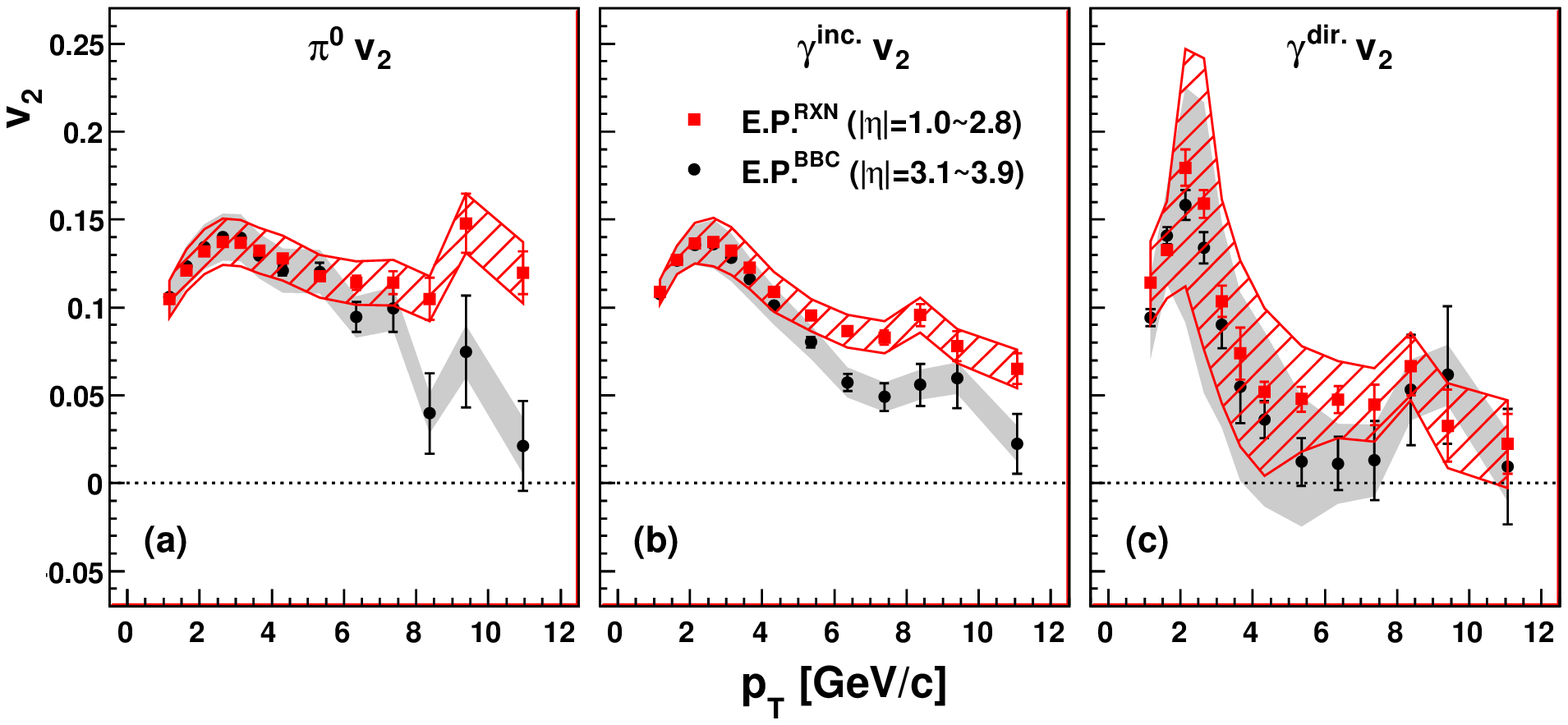}
\includegraphics[width=1.0\linewidth]{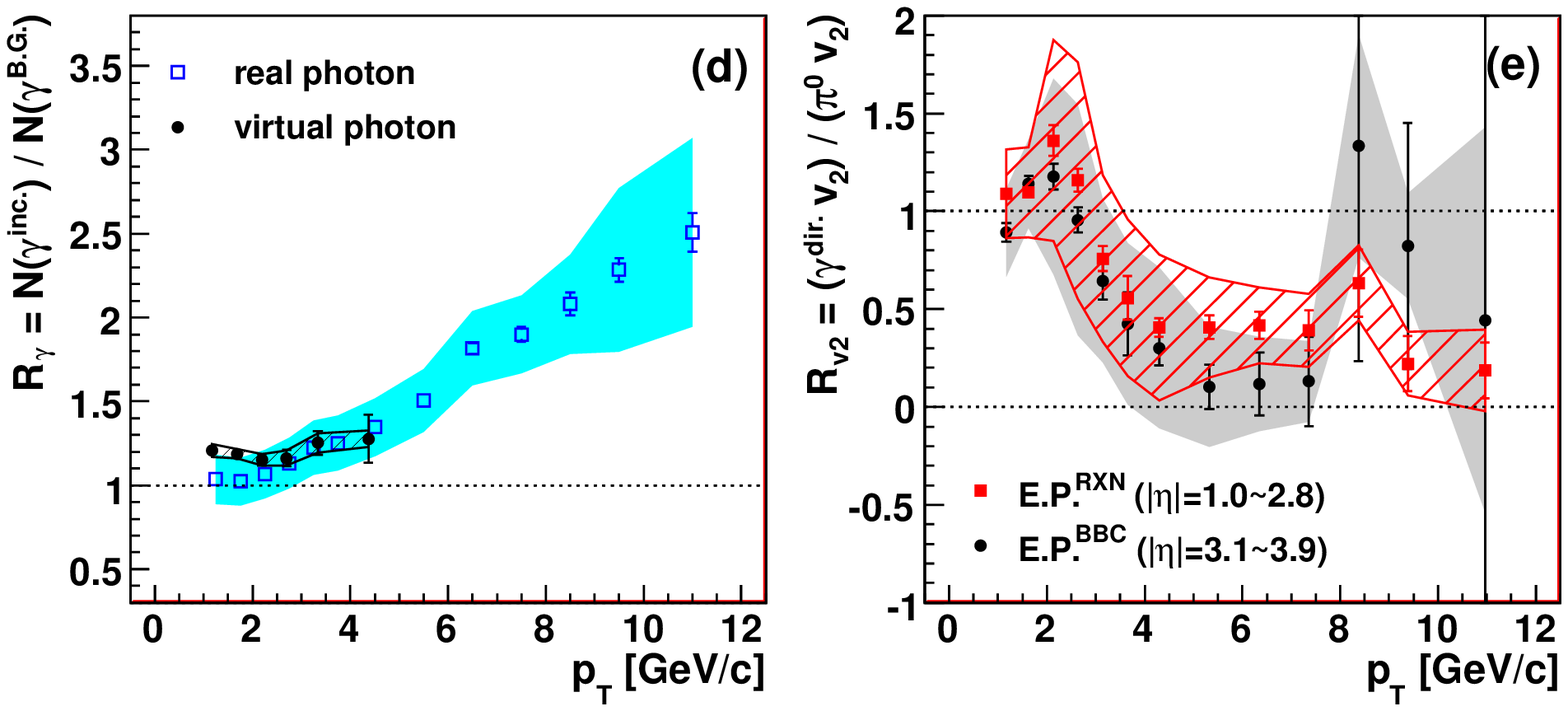}
\caption{\label{fig:fig1} (Color online)
(a,b,c)  \vtwo in minimum bias collisions, using two different
reaction plane detectors: (solid black circles) BBC and 
(solid red squares) RXN for 
(a) $\pi^0$, (b) inclusive photon, and (c) direct photon. 
(d) direct photon fraction \rgam for (solid black circles)
virtual photons~\cite{ppg086} and (open blue squares)  
real photons~\cite{ppg042} and (e) ratio of direct photon
to \piz \vtwo for (solid black circles) BBC and
(solid red squares) RXN.  The vertical error bars on each data point 
indicate statistical uncertainties and shaded (gray and cyan) and 
hatched (red) areas around the 
data points indicate sizes of systematic uncertainties.
}
\end{figure}

Sources of systematic uncertainties for representative \pt values are 
listed in Table~\ref{tab:syserr} along with their characterization:
type A means point-by-point uncertainties which are uncorrelated 
with \pt,  type B means uncertainties that are correlated
(with \pt) and type C is the overall normalization uncertainty, moving all
points by the same fraction up or down.  Since the \vtwo measurement
is a relative one (the azimuthal anisotropy is fit without the need
to know the absolute normalization), the \piz and inclusive photon
\vtwo measurements are largely immune
to energy scale uncertainties which are typically the dominant source
of uncertainty in an absolute (invariant yield) measurement.  
The uncertainties on \vtwo  are dominated by the common
uncertainty on determining $\sigma_{\rm RP}$ and by uncertainties on 
particle  identification.
Uncertainties from absolute yields enter indirectly via
the hadron cocktail (normalization)
and more directly at higher \pt (where the real
photon measurement is used) by the $R_{\gamma}(p_T)$ needed to
establish the direct photon \vtwo.  
Note that due to the way $v_2^{\gamma,{\rm dir}}$ is calculated, once 
$R_{\gamma}$ is large, its relative error contributes to the error on
$v_2^{\gamma,{\rm dir}}$ less and less.

Figure~\ref{fig:fig1} shows steps of the analysis using the
minimum bias sample, as well as the differences between results
obtained with BBC and RXN.  The first \vtwo of 
\piz and inclusive photons ($v_2^{\pi^0}$,$v_2^{\gamma,inc}$) 
are measured, as described above (panels (a) and (b)).  Then, using the
$v_2^{\gamma,bg}$ of photons from hadronic decays and the
$R_{\gamma}$ direct photon excess ratio, we derive the
$v_2^{\gamma,dir}$ of direct photons (panel (c)). 
Panel (d) shows the $R_{\gamma}(p_T)$ values from the 
direct photon invariant yield measurements using 
internal conversion~\cite{ppg086} and real~\cite{ppg042}
photons, with their respective uncertainties.  
Panel (e) shows the ratio of $v_2^{\gamma,dir}$/$v_2^{\piz}$.
We observe substantial direct photon flow in the low \pt region (c), 
commensurate with the hadron flow itself (e).
However, in contrast to hadrons, the direct photon \vtwo rapidly
decreases with \pt; and starting with 5~GeV/$c$ and above, it is 
consistent with zero (c).  The rapid transition from high direct
photon flow at 3~GeV/$c$ to zero flow at 5~GeV/$c$ is also 
demonstrated on panel (e), since the \piz \vtwo changes little in this
region~\cite{ppg092}. 

\begin{figure}[thb]
\includegraphics[width=1.0\linewidth]{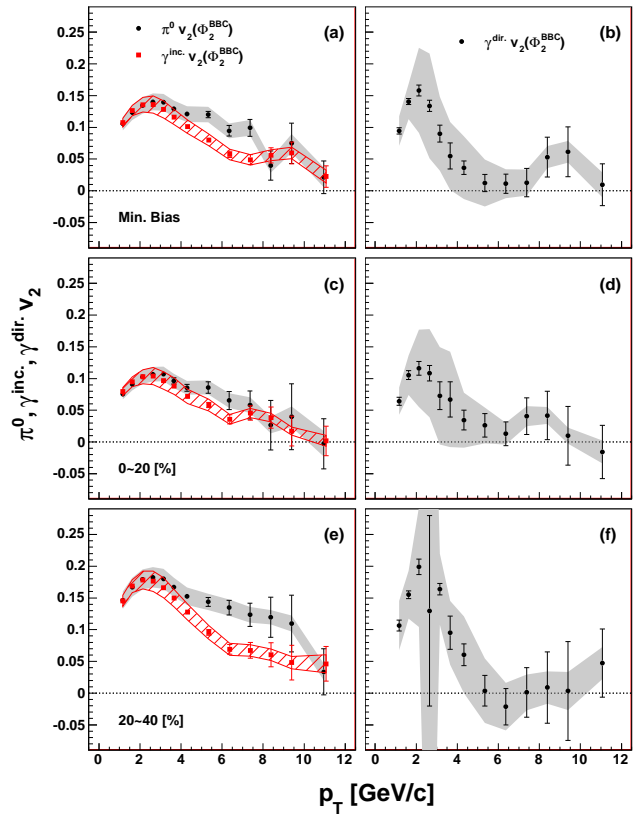}
\caption{\label{fig:fig2} (Color online)
(a,c,e) Centrality dependence of \vtwo for (solid black circles)
\piz, (solid red squares) inclusive photons, and (b,d,f) (solid black 
circles) direct photons measured with the BBC detector for (a,b) minimum 
bias (c,d) 0-20\% centrality, and (e,f) 20-40\% centrality.
For (b,d,f) the direct photon fraction is taken from~\cite{ppg086} up to 
4~GeV/$c$ and from~\cite{ppg042} for higher \pt.
The vertical error bars on each data point indicate statistical 
uncertainties and the shaded (gray) and hatched (red) areas around the 
data points indicate sizes of systematic uncertainties.
}
\end{figure}

A major issue in any azimuthal asymmetry measurement is the potential 
bias from where in pseudorapidity the (event-by-event) reaction plane 
is measured.  
At low \pt -- where multiplicities are high and particle production is
dominated by the bulk with genuine hydrodynamic behavior --
there is no difference between the flow derived with BBC and RXN.
However, at higher \pt we observe that the \vtwo values using BBC and RXN
diverge, particularly for \piz (panel (a) in Fig.~\ref{fig:fig1}), 
less for inclusive photons.  For direct photons (panel (c)) the two
results are apparently consistent within their {\it total} errors, 
including
the error $\delta R_{\gamma}/R_{\gamma}$ (see Table~\ref{tab:syserr})
but it should be noted that $R_{\gamma}$
is a common correction factor in the \vtwo measurements with both
reaction plane detectors.

Event  substructure not related to bulk properties and expansion 
-- most notably jets -- 
can bias the reaction plane measurement, particularly at higher \pt 
and lower multiplicity. 
Observation of a high \pt particle practically 
guarantees the presence of a jet, which in turn modifies the event 
structure over a large \h range.  The bias on the true event plane 
(with the bulk as its origin) is stronger if the overall multiplicity 
is small and if the \h gap between the central arm (where \vtwo is 
measured) and the reaction plane detector is reduced.
The bias in Fig.~\ref{fig:fig1} is largest for \piz, since 
high \pt hadrons are always jet fragments. 
Inclusive photons are a mixture of hadron decay photons, inheriting
the bias seen in \piz and the mostly unbiased direct photons,
therefore, the difference between BBC and RXN is smaller.
Finally, the bias is smallest (but nonzero) for direct photons, 
of which only a relatively small fraction (jet fragmentation photons)
exhibit bias.

Figure~\ref{fig:fig2} shows $v_2$ for minimum bias and two 
centralities as a function of transverse momentum for \piz, 
inclusive and direct photons.  For reaction plane determination the
BBC is used 
because it is farthest from midrapidity where \vtwo is measured.
Despite the fact that there
is a significant direct (thermal) photon yield at low
\pt~\cite{ppg086}, the \piz and inclusive photon \vtwo is virtually 
identical there.  Note that the surprisingly large inclusive photon
\vtwo is confirmed by the (so far preliminary) results with a completely
different analysis technique~\cite{petti}.
For direct photons at low \pt we observe a pronounced positive \vtwo 
signal, increasing with decreasing centrality and comparable 
to the \piz flow, but then rapidly going toward 
zero at 5-6~GeV/$c$.  Qualitatively this shape agrees with the
prediction for very early thermalization times, 0.2-0.4~fm/$c$
in~\cite{srivastava}, 0.2~fm/$c$ and vanishing viscosity
in~\cite{gale}, but both models severely underestimate the magnitude 
of the \vtwo.  The model in~\cite{liu2009} combines somewhat later
thermalization time (0.6~fm/$c$) with partial chemical equilibrium in
the hadronic phase, reproducing the shape, but missing the magnitude of
the observed \vtwo at low \pt.
While such large direct photon \vtwo in principle could be attributed to a
dominant production mechanism at the later stage when bulk flow is 
already developed, simultaneously explaining the large values of \vtwo at
$\sim$2~GeV/$c$ and its vanishing above 5~GeV/$c$ remains a
challenge to current theories.


\begin{figure}[thb]
\includegraphics[width=1.0\linewidth]{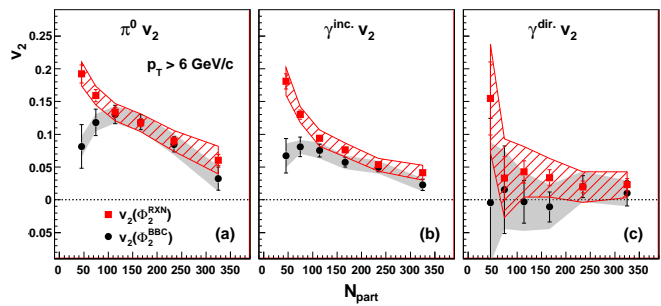}
\caption{\label{fig:Fig3}
High \pt ($p_T>6$~GeV/$c$) integrated \vtwo vs \Npart for (a) \piz,
(b) inclusive photon, and (c) direct photon.  
Results are shown with both reaction plane detectors:
(solid black circles) BBC and (solid red squares) RXN.
Each point represents a 10\% wide centrality bin from 60--0\%.
The vertical error bars on each data point indicate statistical 
uncertainties and the shaded (gray) and hatched (red) areas around the 
data points indicate sizes of systematic uncertainties.
}
\end{figure}

Figure~\ref{fig:Fig3} shows the high \pt
integrated \vtwo ($p_T>6$~GeV/$c$) for \piz 
and photons (inclusive and direct) as a function of centrality.  
The low \Npart behavior is strongly influenced by the
location in pseudorapidity 
of the reaction plane detector.
The \piz \vtwo is comparable to other
hadrons and is higher than the inclusive photon \vtwo, which is
diluted 
by direct photons.  
The two direct photon \vtwo measurements 
(panel (c)) are
consistent with zero (and each other) at all centralities
within their total systematic errors.
While zero \vtwo would
be expected if initial hard scattering is the dominant (sole considered)
source of photons, it should be pointed out that
the typical contribution from jet-conversion only would
be $v_2 \sim -0.02$ and from fragmentation $v_2 \leq 0.01$ weighted
with the fraction of photons coming from these specific
processes~\cite{turbide,gale}. 
Currently the experiment is not sensitive to their
negative/positive contributions to \vtwo.

In conclusion, PHENIX has measured  \vtwo of \piz, inclusive
and direct photons in the $1<p_T<13$~GeV/$c$ range 
for minimum bias and selected centralities  in 
\sqsn = 200~GeV Au$+$Au collisions. 
At higher \pt
($>6$~GeV/$c$) the direct photon \vtwo is consistent with zero at all
centralities, as expected if the dominant source of photon production
is initial hard scattering.
However, the experimental uncertainties are currently
about a factor of 2 higher than the predicted (small) 
positive and negative
contributions from fragmentation and jet conversion photons,
respectively. 
In the thermal region ($p_T<4$~GeV/$c$),
a positive direct photon \vtwo is observed which is
comparable in magnitude to the \piz \vtwo and consistent with
early thermalization times and low viscosity, but its magnitude is much
larger than current theories predict.



We thank the staff of the Collider-Accelerator and Physics
Departments at Brookhaven National Laboratory and the staff of
the other PHENIX participating institutions for their vital
contributions.  We acknowledge support from the 
Office of Nuclear Physics in the
Office of Science of the Department of Energy, the
National Science Foundation, Abilene Christian University
Research Council, Research Foundation of SUNY, and Dean of the
College of Arts and Sciences, Vanderbilt University (U.S.A),
Ministry of Education, Culture, Sports, Science, and Technology
and the Japan Society for the Promotion of Science (Japan),
Conselho Nacional de Desenvolvimento Cient\'{\i}fico e
Tecnol{\'o}gico and Funda\c c{\~a}o de Amparo {\`a} Pesquisa do
Estado de S{\~a}o Paulo (Brazil),
Natural Science Foundation of China (P.~R.~China),
Ministry of Education, Youth and Sports (Czech Republic),
Centre National de la Recherche Scientifique, Commissariat
{\`a} l'{\'E}nergie Atomique, and Institut National de Physique
Nucl{\'e}aire et de Physique des Particules (France),
Ministry of Industry, Science and Tekhnologies,
Bundesministerium f\"ur Bildung und Forschung, Deutscher
Akademischer Austausch Dienst, and Alexander von Humboldt Stiftung (Germany),
Hungarian National Science Fund, OTKA (Hungary), 
Department of Atomic Energy and Department of Science and Technology (India), 
Israel Science Foundation (Israel), 
National Research Foundation and WCU program of the 
Ministry Education Science and Technology (Korea),
Ministry of Education and Science, Russian Academy of Sciences,
Federal Agency of Atomic Energy (Russia),
VR and the Wallenberg Foundation (Sweden), 
the U.S. Civilian Research and Development Foundation for the
Independent States of the Former Soviet Union, 
the US-Hungarian Fulbright Foundation for Educational Exchange,
and the US-Israel Binational Science Foundation.


\end{document}